# Melting of corundum ($\alpha$-$Al_2O_3$) under compression to 8 GPa: A decreasing melting slope

**Vladimir L. Solozhenko**

*LSPM–CNRS, Université Sorbonne Paris Nord, 93430 Villetaneuse, France*

Email: vladimir.solozhenko@univ-paris13.fr

**ABSTRACT:** Melting of corundum ($\alpha$-$Al_2O_3$) has been studied under compression up to 8 GPa using *in situ* electrical resistivity measurements. The melting temperature increases monotonically from $T_0$ = 2317 K at ambient pressure to 3090 K at 7.7 GPa. Fitting the experimental data to the Simon-Glatzel equation yields the empirical parameters $a$ = 5.60 GPa and $c$ = 0.334, providing a closed-form analytical description of the entire melting boundary. Most significantly, the melting slope, $dT_m/dp$, decreases continuously from 138 K/GPa at ambient pressure down to 78 K/GPa at 7.7 GPa, confirming unequivocal negative curvature of the melting curve ($d^2T_m/dp^2 < 0$). This phenomenon is consistent with the preferential densification of the liquid phase and the associated reduction in the volume change upon melting as pressure rises. These findings provide a robust empirical standard for the high-pressure melting of corundum, establishing essential constraints for future *ab initio* simulations, thermodynamic databases, and geophysical models of alumina-rich deep-Earth systems.

## 1. Introduction

Hexagonal (*R*-3*c*) aluminum oxide ($\alpha$-$Al_2O_3$, corundum), a benchmark high-temperature material, finds wide applications as a refractory ceramic, sapphire anvils in high-pressure devices, a thermal insulator, and optical windows in extreme environments. Due to its refractory nature, which is characterized by an exceptionally high melting temperature (2317±4 K at ambient pressure [1]) and chemical stability, the study of its melting under pressure is a challenging task and requires careful assessment. Such an assessment is crucial for interpreting high-energy-density shock experiments and modeling the thermal regime of the deep Earth's mantle.

Despite the crucial importance of this material, data on melting of $\alpha$-$Al_2O_3$ at pressures up to 10 GPa is scarce and contradictory (Fig. 1). Indeed, to date, only a single experimental study has been conducted [2], yielding a mere three data points below 10 GPa, and consequently, the pressure range of 0-10 GPa, which is of practical importance, remains essentially unconstrained. Theoretical efforts [3-7] have been equally challenged. The simple thermodynamic models [3, 4], while offering conceptual simplicity, suffer from a near-total absence of experimental constraints within this pressure range. Molecular dynamics (MD) simulations have been equally problematic. Conventional MD [5] overestimates the melting temperature of $\alpha$-$Al_2O_3$ by 300 K at ambient pressure and 800 K at 25 GPa [6], while more recent reactive force-field MD simulations [7] predict an ambient-pressure melting point of 1670 K, which is notably 28% lower than the experimental value of 2317 K [1]. First-principles calculations based on the density functional theory [8] hold promise for thermochemical assessment of solid $\alpha$-$Al_2O_3$; however, they cannot serve as a reliable basis for describing the melting behavior of corundum under pressure because the quasi-harmonic approximation neglects the anharmonic effects that dominate in the vicinity of melting.

All these persistent discrepancies, coupled with the near absence of systematic experimental mapping in this pressure range, underscore a fundamental gap in our understanding of corundum melting behavior under modest compression. However, it should be noted that all available literature data indicates a near-linear relationship between melting temperature and pressure in the 0-10 GPa range.

Consequently, it is evident that reliable experimental data are imperative for constructing the melting curve of corundum within the 0-10 GPa range. The goal of the present study is to provide a reliable melting curve of $\alpha$-$Al_2O_3$ at moderate pressures, thereby enhancing its status as a benchmark refractory material in technological applications and geophysics.

## 2. Materials and Methods

Fused $\alpha$-$Al_2O_3$ powder (99.6% purity, particle size < 45 μm) was obtained from Zhengzhou Haixu Abrasives Co. Ltd (China) and used as received. High-pressure experiments in the 2-8 GPa range were performed using a toroid-type high-pressure apparatus equipped with a specially designed high-temperature cell capable of reaching 3500 K [9,10]. The cell was pressure-calibrated at room temperature using the known phase transitions of Bi (2.55 and 7.7 GPa), PbSe (4.2 GPa), and PbTe (5.2 GPa). Temperature calibration under pressure was carried out using well-established reference points, including the melting temperatures of Si, NaCl, CsCl, Pt, Rh, Mo, and the Ni–Mn–C ternary eutectic.

The corundum powder was compacted into pellets and placed in direct contact with a graphite heater inside the cell. The appearance of a liquid phase upon heating at a given pressure was detected in situ by electrical resistance measurements, following the method described previously [11-13]. In a separate set of control experiments, it was confirmed that corundum does not react with graphite over the entire studied pressure–temperature range.

## 3. Results and Discussion

The experimental melting temperatures ($T_m$) obtained between 2.6 and 7.7 GPa are summarized in Table 1 and shown in Fig. 2. The data were fitted to the Simon-Glatzel equation [14]

$$T_m(p) = T_0\left(1+\frac{p}{a}\right)^c$$

with $T_0$ fixed at 2317 K [1]. Non-linear least-squares regression yielded the following values: $a$ = 5.60 GPa, $c$ = 0.334. The resulting melting curve,

$$T_m(p) = 2217\left(1+\frac{p}{5.60}\right)^{0.334}$$

reproduces the experimental data with a root-mean-square deviation of approximately ±10 K. The instantaneous melting slopes ($dT_m/dp$) derived from the Simon-Glatzel fit, exhibit a clear and physically consistent monotonic decrease from 138 K/GPa at 0 GPa to 78 K/GPa at 7.7 GPa (see Table 1). This progressive decrease in $dT_m/dp$ with increasing pressure confirms the negative curvature of the melting curve ($d^2T_m/dp^2 < 0$) across the entire pressure range.

This melting behavior definitively rules out a simple linear approximation of the corundum melting boundary over the studied pressure range. The Simon-Glatzel exponent, $c$ = 0.334, is significantly less than unity, quantitatively confirms the concave-down shape of the melting curve. Physically, the decreasing melting slope can be interpreted using the the Clausius-Clapeyron relation:

$$\frac{dT_m}{dp} = \frac{T_m \Delta V_m}{\Delta H_m}$$

where $\Delta V_m$ and $\Delta H_m$ are the volume and enthalpy changes upon melting, respectively. The positive melting slope indicates that $\Delta V_m > 0$, i.e., the solid is denser than the melt. The observed decrease in $dT_m/dp$ indicates that the ratio $\Delta V_m/\Delta H_m$ decreases with pressure. This is consistent with preferential densification of the liquid phase under compression, reducing the solid–liquid density contrast, and possibly increasing the enthalpy of fusion at higher pressures.

Alternative melting models were considered but found less suitable. A linear model cannot reproduce the pronounced curvature of the experimental data. The Lindemann criterion offers a microscopic rationale for the concave-down trajectory but requires pressure-dependent Grüneisen and Debye parameters that are not well constrained for corundum at high pressures. In contrast, the Simon-Glatzel equation, provides a compact, closed-form, and precise empirical representation of the entire melting boundary, requiring only two adjustable parameters.

These results provide a robust experimental benchmark for the high-pressure melting behavior of $\alpha$-$Al_2O_3$ up to 8 GPa. The determined parameters of the Simon–Glatzel equation ($T_0$ = 2317 K, $a$ = 5.60 GPa, $c$ = 0.334) are essential for thermodynamic databases, *ab initio* simulations, and geophysical models involving alumina-rich phases. The decreasing melting slope further implies that linear extrapolations of the corundum melting curve to higher pressures would result in a significant overestimation of melting temperatures. Extending measurements beyond 10 GPa will be essential to ascertain whether the observed concave-down trend persists or eventually saturates.

## 4. Conclusions

In summary, the melting curve of corundum ($\alpha$-$Al_2O_3$) has been precisely determined up to 8 GPa. The experimental dataset is well-described by the Simon–Glatzel formalism, with $T_0 = 2317$ K, $a = 5.60$ GPa, and $c = 0.334$. The principal finding is the continuous decline in melting slope $dT_m/dp$ from 138 to 78 K/GPa over 7.7 GPa, demonstrating that the solid–liquid density contrast diminishes significantly under compression. This behavior definitively precludes any linear extrapolation of the melting curve of $\alpha$-$Al_2O_3$ and highlights the necessity of incorporating pressure-dependent thermodynamic parameters — specifically, the progressive reduction in the fusion volume ($\Delta V_m$) relative to the enthalpy of fusion ($\Delta H_m$) — when modeling the melting of refractory oxides at high pressures.

These experimental results provide a critical benchmark for computational studies, including density functional theory (DFT) and molecular dynamics (MD), which frequently exhibit significant discrepancies in predicting melting slopes for highly ionic oxides.

Finally, the precise melting boundary established here informs the design of advanced ceramic and thermal barrier coatings that can withstand extreme thermomechanical loads. Furthermore, these data provide essential input for geophysical models of subducting slabs and mantle transition-zone compositions, where alumina-rich phases exert a significant influence on thermal structure and rheology.

## Acknowledgments

The author is grateful to Dr. Vladimir A. Mukhanov for assistance with high-pressure experiments.

### Ethics Statement

Not applicable.

### Informed Consent Statement

Not applicable.

### Data Availability Statement

The data that support the findings of this study are available from the author upon reasonable request.

### Funding Statement

The author received no specific funding for this study.

### Declaration of Competing Interest

The author declares no conflicts of interest.

Table 1 Experimental and fitted melting temperatures and melting slopes of α-$Al_2O_3$.

| Pressure (GPa) | $T_m$ (K) | | $dT_m/dp$ (K/GPa) |
|---|---|---|---|
| | experiment | fit | |
| 0 | 2317 | 2317 | 138 |
| 2.6 | 2635 | 2632 | 107 |
| 3.4 | 2723 | 2715 | 101 |
| 4.2 | 2773 | 2793 | 95 |
| 5.2 | 2893 | 2885 | 89 |
| 6.5 | 3001 | 2997 | 83 |
| 7.7 | 3090 | 3093 | 78 |

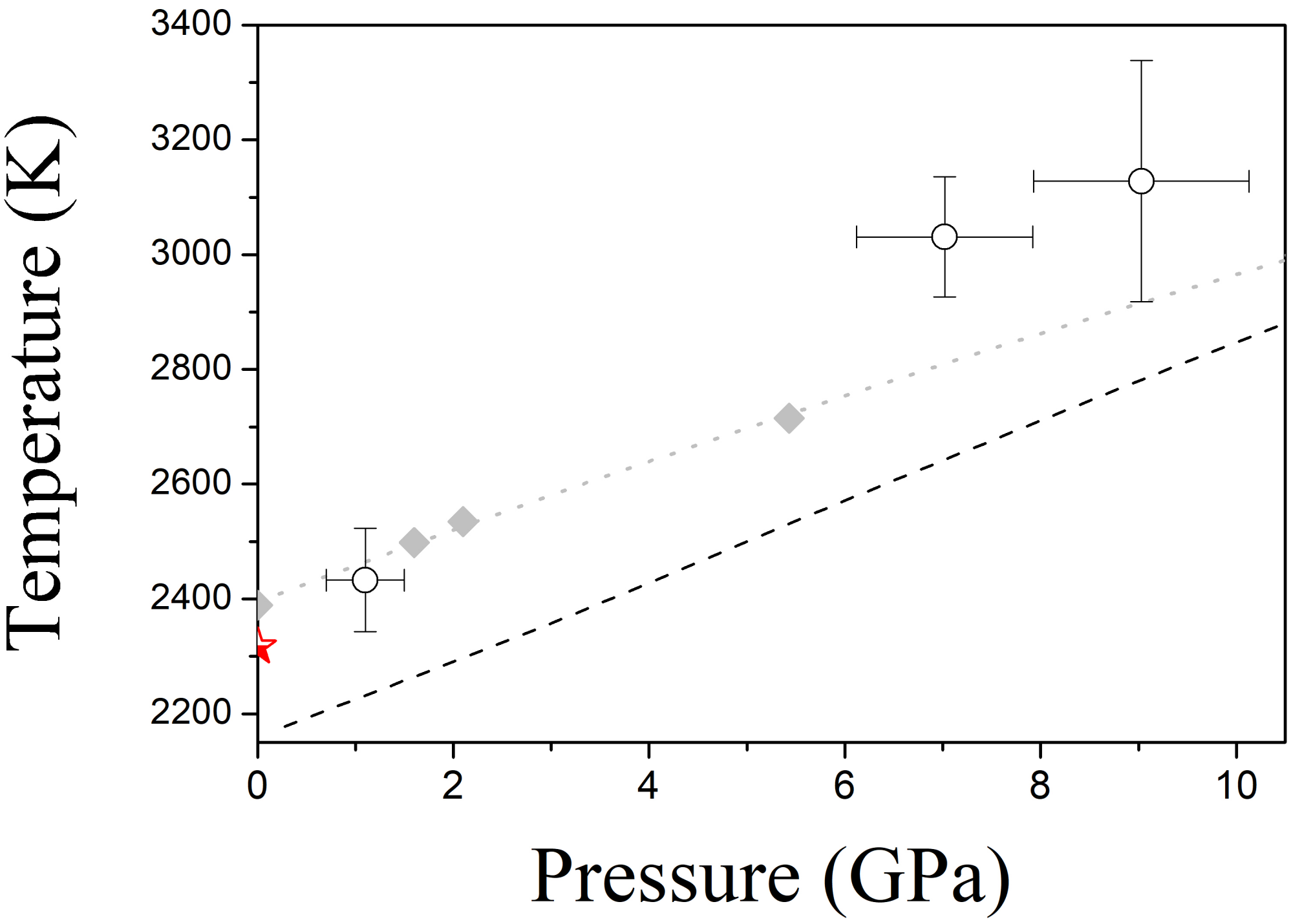

Fig. 1 Literature data on corundum melting in the 0-10 GPa range. The open circles represent the experimental data [2]; the gray filled squares and the dotted line show results from the simple thermodynamic model [3]; and the black dashed line represents results from the molecular-dynamics simulation [5]. The red half-filled star indicates the melting point at ambient pressure [1].

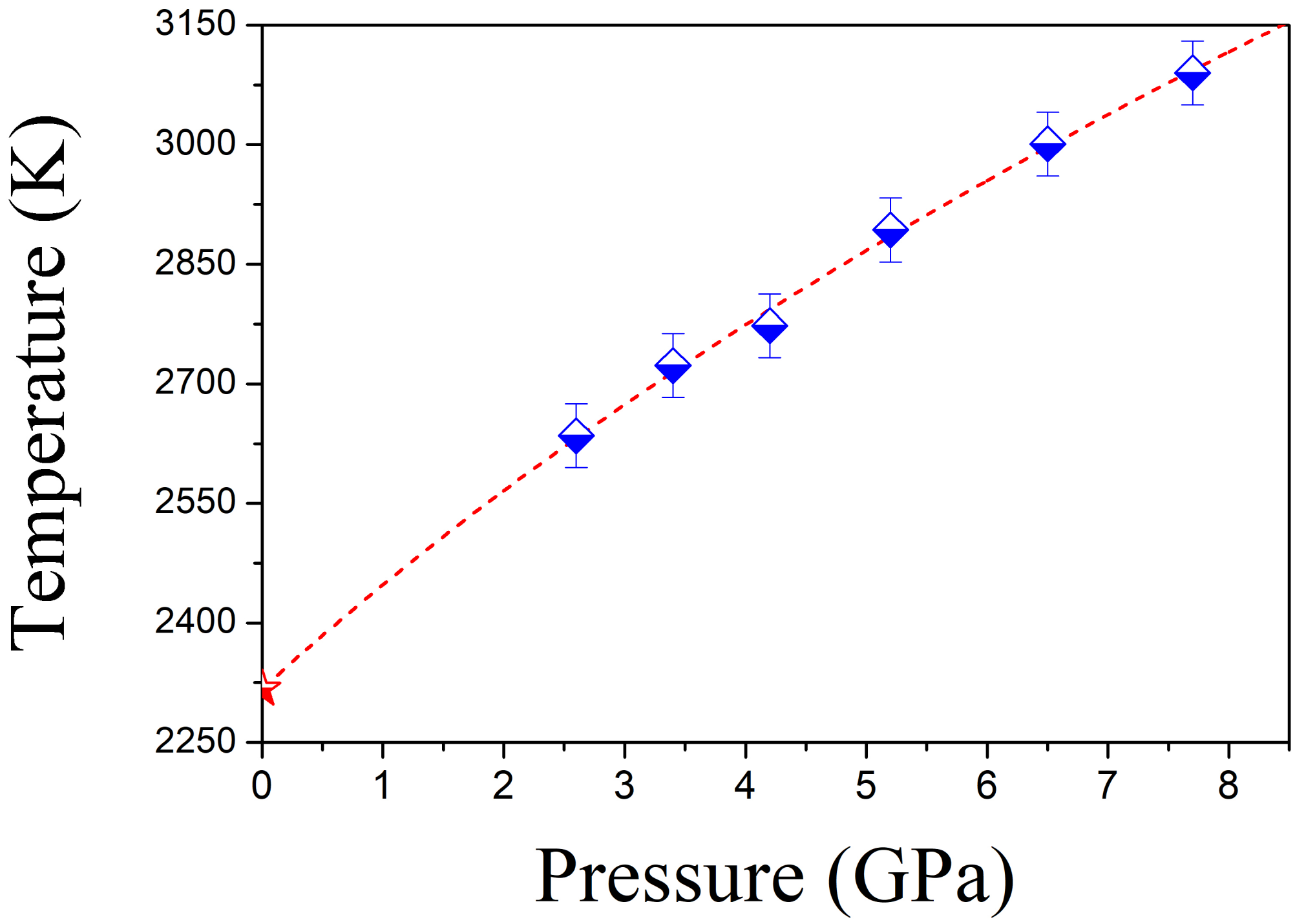

Fig. 2 Experimental melting curve of corundum up to 8 GPa. The blue half-filled diamonds represent the melting temperatures derived from *in situ* electrical resistivity measurements, while the red dashed line shows the corresponding fit to the Simon-Glatzel equation [14]. The red half-filled star denotes the ambient-pressure melting point [1].